\documentclass{PoS}
\usepackage{amsmath}
\usepackage{subcaption}
\usepackage{graphicx}
\DeclareMathOperator{\sinc}{sinc}
\title{Astrophysical measurements with the VERITAS Stellar Intensity Interferometer}
\ShortTitle{Astrophysical measurements with the VERITAS Stellar Intensity Interferometer}
\author{\speaker{Nolan Matthews}$^{1}$ for the VERITAS Collaboration\footnote{http://veritas.sao.arizona.edu}, S. LeBohec $^{1}$ \\
        $^{1}$ Department of Physics and Astronomy, University of Utah, Salt Lake City, UT 84112, USA \\
        E-mail: \email{nolankmatthews@gmail.com}
        }

\abstract{Imaging Atmospheric Cherenkov Telescopes have long been viewed as potential light collectors to be used for  long baseline optical intensity interferometry observations. Intensity interferometry, as implemented with Cherenkov telescopes, is well suited for studying the spatial structure of stars of O/B/A stellar types at short optical wavelengths. Such observations complement those with the current generation of optical amplitude interferometers, which are typically restricted to longer wavelengths. Dedicated intensity interferometry instrumentation has been developed for the VERITAS observatory with engineering tests and observations occurring since October 2018. In this presentation, the first results using two of the VERITAS telescopes, are reported. The system has already been extended to the two additional telescopes, enabling SII observations with all four VERITAS telescopes.}
\FullConference{36th International Cosmic Ray Conference -ICRC2019-\\
		July 24th - August 1st, 2019\\
		Madison, WI, U.S.A.}

\begin{document}

\section{Introduction}
Robert Hanbury-Brown and Richard Q. Twiss showed in the late 1950's that correlations of star-light intensity fluctuations recorded in two different locations are related to the source spatial coherence function and thus probe the size and shape of the source \cite{HBT1957a,HBT1957b}. 
This breakthrough led to the development of the Narrabri Stellar Intensity Interferometer (NSII), which operated from 1963 to 1974 and was used to successfully measure the angular diameters of 32 stars \cite{HBT1974}. Over the past 15 years, there has been a renewal of interest in the technique largely due to the emergence of telescope arrays of large light collectors along with the development of fast detectors with improved quantum efficiency. These factors subsequently allow for a substantial improvement in the limiting magnitude of a modern SII observatory compared to the NSII. SII measurements can be made at short optical wavelengths thus complementing the current generation of modern amplitude interferometers that are generally restricted to longer wavelengths. 

As a result, several groups have pursued SII measurements \cite{aquaeye,weiss}. Recently, there was the report of first measurements of starlight coherence using intensity interferometry on stars other than our sun since the NSII. These measurements were performed with a single telescope \cite{guerin2017}, as well as between a pair of telescopes separated by 15$\,$m \cite{guerin2018}. Broadly speaking, there are two complementary approaches to a modern SII observatory. One is to use optical telescopes allowing precision spectral filtering together with extremely fast detectors with resolution timescales of tens of pico-seconds. The other direction, addressed in this work, is to take advantage of current and future Imaging Air Cherenkov Telescope (IACT) arrays, constructed for high energy gamma-ray astronomy, and outfit them with SII capabilities. The large light collection areas, fast (nano-second resolution) detectors, and layouts of IACT arrays make them extremely compatible for SII measurements \cite{SIIwIACT}. In principle, stars as faint as $m_V \approx 6$ can be studied with SII using IACT observatories \cite{nunez_2012}. The main drawback of currently operating IACTs for SII is the poor optical quality of the reflectors when compared to conventional optical telescopes. Typical IACTs of the Davies-Cotton optical design exhibit a point-spread-function (PSF) on the order of 0.1$^\circ $. Ultimately, the fundamental limiting magnitude of potential SII targets is set by the PSF due to the contribution of the night sky background (NSB) flux \cite{janvida_2013}. Nevertheless, an SII system on an IACT array with modern instrumentation offers the possibility to characterize stars over 3 magnitudes dimmer than what was possible with the NSII, comparable to the current limits of modern amplitude interferometers operating in the visible bands \cite{Mourard_2009}. 

A number of IACT observatories are currently operating that could be used for SII observations including MAGIC \cite{MAGIC}, HESS \cite{HESS}, and VERITAS \cite{veritas_upgrade}. The next generation IACT observatory, the Cherenkov Telescope Array (CTA) \cite{ctaconcept}, is currently under construction and will be composed of two arrays, one located in each hemisphere. The northern site, which will be located in La Palma, Spain, is planned for 19 telescopes with diameters of 12\,m and 23\,m dispersed over an area that is approximately 800\,m in diameter. The southern site, which will be located in Paranal, Chile, is expected to have 99 telescopes dispersed over an area of about 2.4 km in diameter \cite{ctalayout}. The telescopes in the southern site include the same large and medium sized telescopes as the northern site, and are completed by a more extended array of smaller 4$\,$m diameter telescopes. Furthermore, CTA may incorporate a new type of IACT based on the Schwarzschild-Couder optical design which has a significantly smaller PSF \cite{Vassiliev2007}. Numerous science cases are possible using CTA as an SII observatory including, for example, the study of stellar limb-darkening, resolving bright and cool star-spots \cite{siiCTA}, characterizing oblate stars due to rapid rotation, and potentially model-independent imaging due to the dense uv-plane coverage \cite{nunez1, nunez2}.      

\section{SII with the VERITAS telescopes}

IACT observatories currently in operation can be used as prototyping platforms for future SII observations with large arrays and at the same time allow scientifically useful astrophysical application of SII. There is such a program ongoing with VERITAS, which is an array of four identical 12$\,$m diameter Davies-Cotton telescopes, located in Amado, Arizona, in the Fred Lawrence Whipple Observatory. Over the summer of 2018, an SII system was constructed for the VERITAS telescopes, based on prior instrumentation developed in the laboratory and tested at the StarBase-Utah Observatory \cite{matthews1,starbase1}. More details of the instrumentation are described in  \cite{kieda_icrc2019}. Engineering tests began during fall 2018, with on-sky observations using two of the telescopes starting in December 2018. 

An SII system allows for the measurement of the second order coherence function, $g^{(2)} (\tau)$. For a randomly polarized thermal light source, in the case where the electronic bandwidth $\Delta f$ is much less than the optical bandwidth $\Delta \nu$,  $g^{(2)} (\tau)$ measures the interferometric squared visibility $|V(\mathbf{r})|^2$ at zero time-lag

\begin{equation}
    g^{(2)} (\mathbf{r}, \tau=0) = \frac{\langle I_1 I_2 (\mathbf{r}) \rangle}{\langle I_1 \rangle \langle I_2 \rangle} = 1 + \frac{\Delta f}{\Delta \nu}|V(\mathbf{r})|^2
\end{equation}
where $\mathbf{r} = \mathbf{r_1} - \mathbf{r_2}$ is the projected separation between the telescopes that measure the respective light intensities $I_1$ and $I_2$ as a function of time. Here, $\tau$ is defined as the time-lag between the telescopes after intrinsic cabling delays and geometrical optical path delays $\tau_{OPD}$ have been taken into account. The complex visibility $V(\mathbf{r})$ is the Fourier transform of the source brightness distribution in the sky. Consider a star that can be described as a disc of uniform intensity up to an angular diameter $\theta$. The visibility as a function of the telescope separation then takes the form of an airy-disk profile, 

\begin{equation}
    V(\mathbf{r}) = 2 \frac{J_1 (\pi \theta \mathbf{r}/\lambda)}{\pi \theta \mathbf{r}/\lambda}
\label{eqn:unif_disk_visb}
\end{equation}
where $J_1$ is the first order Bessel function of the first kind. For a given wavelength $\lambda$, the visibility for larger sources will fall off quicker than for smaller ones, and reaches a first zero at a baseline of $r \approx 1.22 \lambda / \theta$, often referred to as the resolving baseline. 

The first two VERITAS telescopes outfitted with SII capabilities (T3 and T4) are separated by a North-South, East-West, and Up-Down baseline of 65.3$\,$m, 48.8$\,$m, and 2.8$\,$m, respectively. This pair was specifically chosen for the first tests because the radial separation of 81.5$\,$m (when viewed from zenith) is the smallest among all telescope pairs, and from Equation \ref{eqn:unif_disk_visb} maximizes the visibility. Note however, that since $r$ is the projected separation between the telescopes as viewed from the star, its value will change as the star is tracked through the sky. This effect makes it possible to measure the squared visibility over a range of baselines $r$ with a single fixed telescope pair, a process known as uv-plane synthesis \cite{segransan}. At the nominal observing wavelength of $\lambda = 415\,$nm, the maximum baseline between the T3/T4 telescopes becomes the resolving baseline for sources that are $\sim 1.3\,$mas in angular diameter. Stars with diameters less than 1.3\,mas bright enough for their spatial coherence to be detected within less than one hour of observation were identified as good candidates for first tests. The star $\gamma$ Ori (Bellatrix), with a B band magnitude of $m_B$ = 1.42, and a measured uniform disk diameter of $\theta_{UD} = 0.701 \pm 0.005\,$mas \cite{chara1}, fit these criteria and was chosen for initial observations.
\begin{figure}[t]
    \centering
    \includegraphics[width=0.8\textwidth]{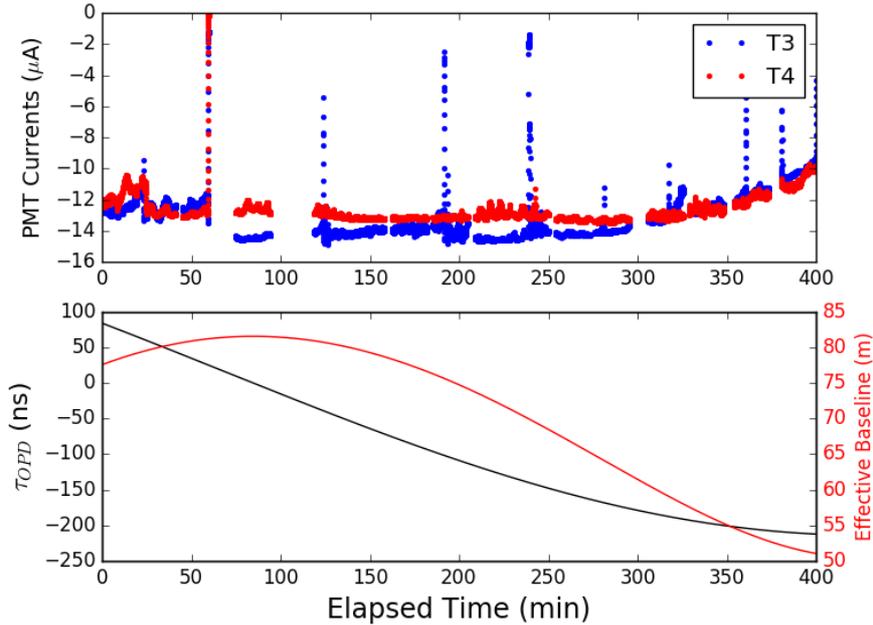}
    \caption{The top panel displays the PMT currents averaged over 1 second cycles throughout the duration of the night for both telescopes. Note more negative values correspond to higher current. The short periods of time where significant current is lost is associated to tracking corrections during which the star is momentarily off-centered in the SII optics. A loss of current is seen at the end of the night and is attributed to atmospheric absorption of starlight under large zenith angles. The bottom panel displays both the optical path delay between telescopes (black) and the projected radial baseline (red).}
    \label{fig:gamoridiags}
\end{figure}

Observations of $\gamma$ Ori presented herein were conducted on the night of January 22nd, 2019. Figure \ref{fig:gamoridiags} shows the evolution the one second averaged PMT currents, the geometrical optical path delay, and the projected telescope separation.  
Data was continuously recorded throughout the night in a series of typically twenty minute runs. At the end of the night, the time-series data is passed to an FPGA correlator. The correlator calculates a series of correlograms corresponding to 1 second accumulation cycles, which each cover a relative time-lag range from -128$\,$ns to +124$\,$ns in steps of the sampling period. The mean optical path delay $\tau_{OPD}$ is estimated for each accumulation cycle from the relative baseline and source direction. The optical time delay is then corrected by the intrinsic cabling relative time delays. 
A data quality cut is then applied to the data. This cut removes each one second cycle contaminated by any high frequency noise that may cause unwanted spurious correlations. Fortunately, the noise pickup is well localized within a narrow frequency band, allowing identification by means of a Fourier transform of the correlograms. Correlograms displaying an excess power at the noise frequency are discarded.  All remaining one second correlograms are then averaged together. 

\begin{figure}[!tp]%
    \centering
    \begin{subfigure}{0.49\textwidth}
        \includegraphics[width=\linewidth]{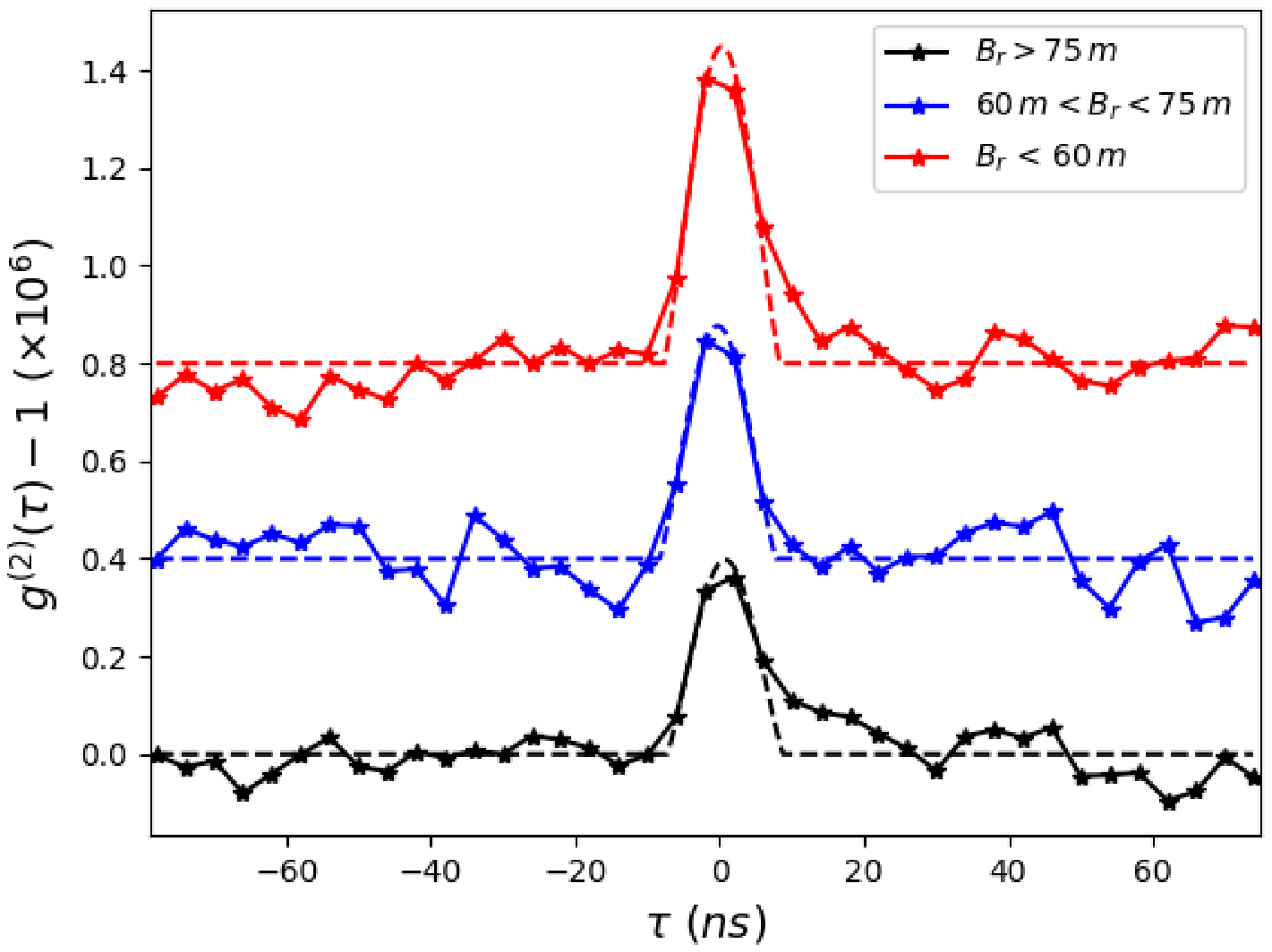} 
    \end{subfigure}
    \begin{subfigure}{0.49\textwidth}
        \includegraphics[width=\linewidth]{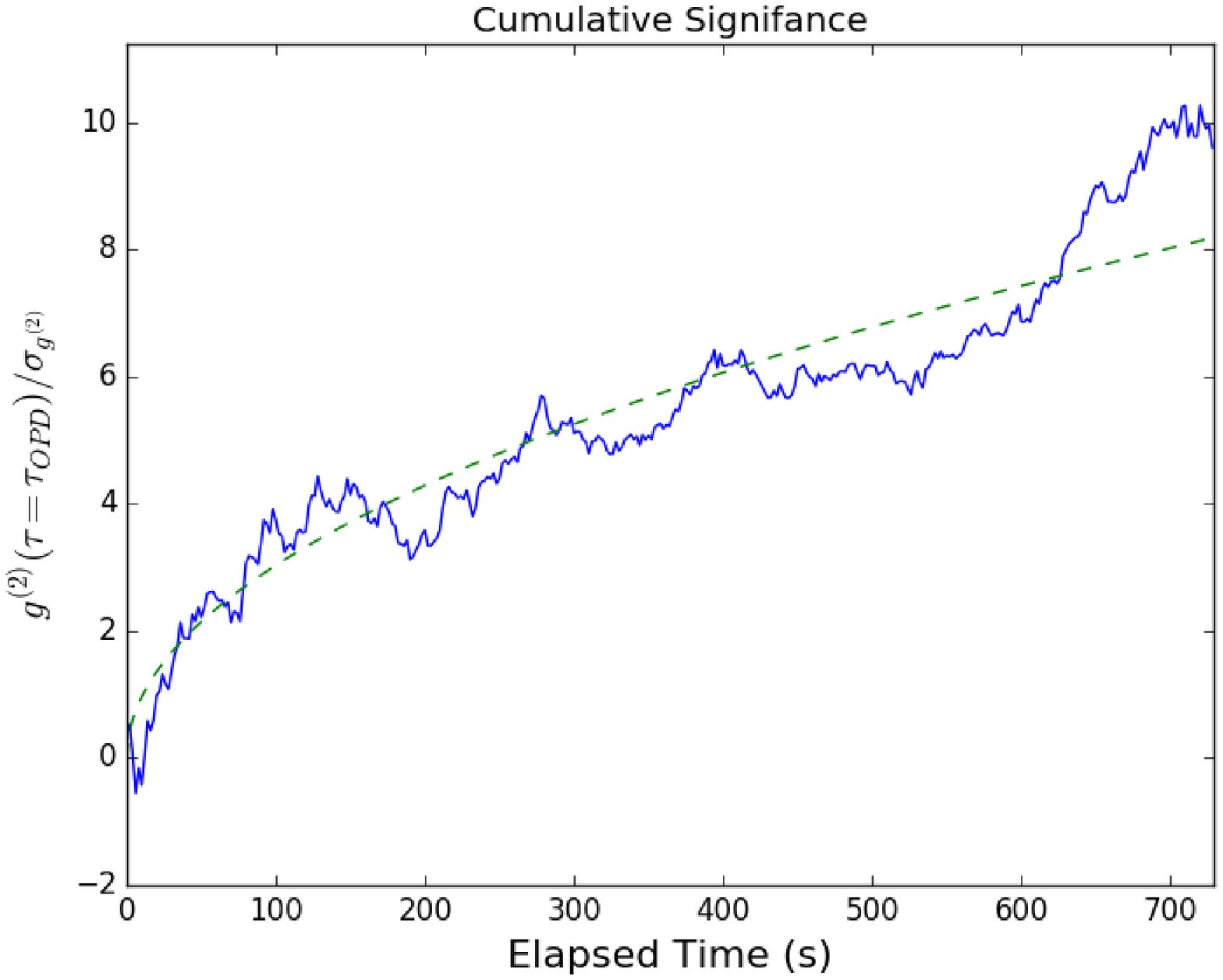}
    \end{subfigure}
    \caption{Results from observations of $\gamma$ Ori on UTC 2019-01-23. The plot on the left shows the averaged time-corrected correlation binned in different baseline ranges of < 60 m (red), between 60 and 75 m (blue), and greater than 75 m (black), with each bin corresponding to a total of 45, 42, and 85 minutes of integrated data, respectively, after noise-cuts are applied. A clear excess is seen at time-lags corresponding to the optical path delay between telescopes and is attributed to the spatial coherence seen between the two telescopes. In addition, as the baseline decreases, the excess correlation, which is an effective measure of the squared visibility, increases. The dotted lines are fits to the data and is described in the text. The plot on the right shows the cumulative significance of over one of the runs taken during the night after noise-cuts have been applied. The dotted line is a fit to the data for a function of the form SNR $= A \sqrt{t}$, where A is left as a free parameter.}%
    \label{fig:gamoriresults}%
\end{figure}

Figure \ref{fig:gamoriresults} shows the results from $\gamma$ Ori observations. Data taking began at 2019-01-22 19:12:56 MST when the source at an elevation of 47$^{\circ}$ before culmination and continued until the source dropped to 20$^\circ $ in elevation at 01:52:21 MST. The panel on the left displays the integrated correlogram after time-delay corrections and noise cuts are applied. The peak at zero time-lag is the signal associated with the spatial coherence of the source. The peak is fit to a function of the form $g^{(2)} (\tau) \propto \sinc{(\pi \Delta f \tau})$ for $|\tau| < 1/\Delta f$ which is derived under the assumption that each channel displays a rectangular electronic bandwidth of width $\Delta f$ \cite{hbtbook}. For the sampling rate of 250\,MS/s used in these observations, the corresponding Nyquist bandwidth is $\Delta f = 125$\,MHz. This function is fit to the data including free parameters for the peak time-lag and normalization. 
The right panel shows the ratio of the correlation measured at zero time-lag relative to the root-mean-square fluctuations of the correlation in a correlogram off-region where no signal is expected. The correlogram off-region is defined by any time-lags where $|\tau| > 2/\Delta f$. The cumulative signal to noise (SNR) ratio of the $g^{(2)}$ peak can be written as, SNR $\propto \eta |V(r)|^2 \sqrt{t}$ where $\eta$ is the measured spectral flux. Overall, the cumulative significance grows as expected, indicating that the system is well-behaved.                  

\section{Conclusions}

The results here demonstrate the feasibility of SII measurements with IACT arrays and the capabilities of the first "off-line" optical interferometer, where the data from each telescope is streamed to disk with correlations produced post-observation. Furthermore, these observations were taken near full-moon conditions when VERITAS does not operate, thus highlighting the ability of IACT arrays to perform SII measurements without impacting the primary gamma-ray astronomy objectives of the facilities.   

The current SII system is being upgraded in several ways. The sampling resolution time can be increased to improve the sensitivity until it becomes shorter than the detector time-resolution or the arrival time-spread of photons in the focal plane resulting from the Davies-Cotton design of the telescope reflecting surface. The current sampling rate of 250 MS/s can be doubled to 500 MS/s to offer up to a $\sqrt{2}$ improvement of the signal-to-noise ratio of the correlation. Currently, the spurious correlations render $\sim$ 30$\%$ to 50$\%$ of the data unusable. Hardware improvements designed to reject or suppress the high frequency noise would then allow for a significant improvement in the observing efficiency. Although an exact solution is currently unknown, the data corruption can be minimized by improved pre-amplifiers that are less susceptible  to the noise, notch filters centered about the noise frequency, or, if the noise can be well-characterized, by removing the noise in software. Optical filters in separate spectral bands would allow broader uv-plane coverage and also could be chosen in order to maximize the spectral flux for a given target due to differences in spectral types. Due to the f/1 focal ratio and large PSF, it is challenging to measure correlations over multiple spectral channels simultaneously with the VERITAS telescopes. However, measurements over multiple channels can be performed serially in time with, for example, the use of a filter wheel. In principle, the pass-band of the filters can be chosen from a lower limit of $\sim$ 350\,nm set by atmospheric attenuation to greater than $700\,$nm where the mirror reflectivity is still greater than $\sim$ 70\%. For wavelengths greater than $\sim$ 450$\,$nm the quantum efficiency of the current detectors degrades rapidly $\cite{nepomuk_icrc2011}$, and so other detectors would be required for observations at these wavelengths. 

Further steps include estimating the stellar angular diameter by measuring the squared visibility as a function of the projected baseline at the time of observation. Once the stellar diameters are verified, a variety of astrophysical targets can be studied using all four of the VERITAS telescopes by taking advantage of the greater uv-plane coverage. The array is particularly well suited for the characterization of rapidly rotating stars \cite{belle_rr} by resolving the equatorial to polar radii ratio. Good candidates include target stars such as $\eta$ UMa, $\gamma$ Cas, and $\eta$ Tau. Observations of binary systems could provide constraints on the orbital parameters but in addition, observations with VERITAS could be uniquely suited to study the properties of the individual stars. A great candidate is $\alpha$ Vir which was first studied in interferometry using the NSII \cite{herbisonevans_1971} and also more recently with optical amplitude interferometery \cite{aufdenberg_2006}.

\section{Acknowledgements}

This research is supported by grants from the U.S. Department of Energy Office of Science, the U.S. National Science Foundation and the Smithsonian Institution, and by NSERC in Canada. This research used resources provided by the Open Science Grid, which is supported by the National Science Foundation and the U.S. Department of Energy's Office of Science, and resources of the National Energy Research Scientific Computing Center (NERSC), a U.S. Department of Energy Office of Science User Facility operated under Contract No. DE-AC02-05CH11231. The authors gratefully acknowledge support from NSF Grant No. AST 1806262 for the fabrication and commissioning of the  VERITAS-SII instrumentation. We acknowledge the excellent work of the technical support staff at the Fred Lawrence Whipple Observatory and at the collaborating institutions in the construction and operation of the instrument.

\end{document}